\documentclass{JHEP3}
\usepackage{amsmath}
\usepackage[numbers,sort&compress]{natbib}
\usepackage{graphicx}
\usepackage{url}
\usepackage[bf,small]{caption2}
\setcaptionwidth{.9\textwidth}
\makeatletter
\renewcommand\@ENVwarn[1]{}
\makeatother

\title{Mesons versus quasi-normal modes: undercooling\\ and overheating}

\author{Angel Paredes{}$^1$, Kasper Peeters{}$^1$ and Marija Zamaklar{}$^2$\\
\llap{{}$^1$}Institute for Theoretical Physics, Utrecht University, P.O.~Box 80.195,
3508 TD Utrecht, The Netherlands.\\
\llap{{}$^2$}Department of Mathematical Sciences,
Durham University,
South Road,
Durham DH1 3LE, United Kingdom.\\
~\\
\email{a.paredesgalan@uu.nl}\\
\email{kasper.peeters@aei.mpg.de}\\
\email{marija.zamaklar@durham.ac.uk}}

\abstract{In holographic models of gauge theories with matter, there
  generically exists a first order phase transition in which mesons
  dissociate. We perform a careful analysis of the meson and
  quasi-particle spectra in the overheated resp.~undercooled regimes
  close to the junction of the two phases. We show that all overheated
  finite meson masses eventually become infinitely degenerate, which
  implies that any connection to the quasi-particle spectrum is
  more subtle than previously suggested. For the Sakai-Sugimoto model
  no smooth connection between these spectra is possible.}

\keywords{AdS/CFT, mesons, phase transitions}
\preprint{\small ITP-UU-08/04, SPIN-08/04, DCPT-08/11}

\begin{document}
\section{Introduction}

Holographic models for gauge theories with matter exhibit a rather
generic three-phase structure. For low temperatures, the background is
a regular geometry, representing the confined phase of the gauge
theory. At sufficiently high temperature, the background undergoes a
Hawking-Page transition to a black brane geometry. In the dual gauge
theory, this transition corresponds to the deconfinement of
gluons. Mesons still remain bound, as long as the brane representing
the matter degrees of freedom does not intersect the background
horizon. When this finally happens, a second transition occurs, to a
phase in which mesons melt and the gauge theory is fully deconfined.
We will call these three phases the low, intermediate and high
temperature phase respectively.\footnote{There is by now a large list
  of references dealing with this three-phase structure; seminal
  papers are~\cite{Babington:2003vm} for the D3/D7 system,
  \cite{Kruczenski:2003uq} for the D4/D6 case and
  \cite{Aharony:2006da} for the D4/D8 model. In
  special situations the intermediate phase can be absent, but we will
  not discuss those cases here.}

After the mesons have melted, one is left with a strongly-coupled
quark-gluon fluid. This fluid exhibits a discrete spectrum of unstable
excitations~\cite{Hoyos:2006gb}, which are described by quasi-normal
modes on the probe brane. It has been
suggested~\cite{Myers:2007we,Erdmenger:2007jab} that the thermal pole
masses of mesons before the transition should be related to the real
parts of the quasi-normal modes after the transition. It would then be
possible to follow the stable mesons of the intermediate temperature
phase into the high-temperature phase, where they become unstable
quasi-particles with finite decay width. The fact that the widths are
of order one is consistent with the expected behaviour of meson decay
widths in the deconfined phase.\footnote{In the
  intermediate-temperature phase the decay of a meson to a
  quark-antiquark pair would also be of order one, but because mesons
  are anomalously light (of order~$2m_q/\sqrt{\lambda}$ where~$m_q$ is
  the constituent quark mass), there is no phase space available for
  their decay. Meson to two-meson decays~\cite{Peeters:2005fq} are of
  order~$1/N_c$.}  Moreover, lattice results indicate that
finite-width mesonic resonances indeed survive above the deconfinement
temperature~(see e.g.~\cite{deForcrand:2000jx,Oktay:2007xb}).

While such a connection between stable and unstable meson modes thus
seems reasonable, it so far relies on numerical evidence. Other known
facts about the generic thermal behaviour of meson masses suggest that
the connection might be more subtle. In the intermediate temperature
phase, the pole masses of mesons are known to decrease as the
temperature goes up~\cite{Peeters:2006iu}. This shows qualitative
resemblance with lattice QCD
studies~\cite{Karsch:1999vy,Gottlieb:1996ae}. In QCD proper, chiral
partners such as the~$\rho$ and $a_1$ meson eventually become
degenerate when chiral symmetry is restored. In contrast, all
holographic models go through the first-order meson melting phase
transition before such degeneracies can be observed. However, the
graphs of the thermal behaviour of the masses~\cite{Peeters:2006iu}
suggest that a degeneracy may still occur when the system is
overheated. If such a degeneracy occurs, then an identification of the
discrete spectrum before and after the transition becomes more subtle
(we will indeed see that this happens).

A further problem occurs in the Sakai-Sugimoto model with zero bare
quark mass. The embeddings after the transition are all equivalent,
which implies that the quasi-normal mode spectrum after the transition
is fixed, i.e.~it never becomes degenerate. If the meson spectrum, on
the other hand, does become degenerate at the junction, then no smooth
identification seems possible. In addition,\footnote{We thank Ofer
  Aharony for emphasising this point to us.} mesons of this model
transform under the adjoint of the diagonal~$SU(N_f)$, while the
quasi-normal modes sit in the adjoint of~$SU(N_f)_L$ or
$SU(N_f)_R$. This again makes a straightforward identification of the
spectra unlikely.
\medskip

In the present paper we therefore try to answer two key questions:
what is the precise behaviour of meson pole masses as the critical
embedding is approached from below the transition temperature, and
does this leave any room for a connection to the real parts of the
frequencies of the quasi-normal modes just above the critical
temperature? We will restrict ourselves to an analysis of the spectrum
of \emph{vector} excitations close to the critical embedding, and
comment only briefly on a similar analysis for scalar excitations.
For the sake of simplicity, we will also only consider modes
homogeneous in the internal spheres; generalisation to other cases
should be straightforward.

We will first provide numerical evidence which indicates that the
potential of the associated Schr\"odinger problem takes a particularly
simple form as the critical embedding is approached. Inspired by these
numerical results, we then present an analytic procedure to compute
the spectrum close to either side of the phase transition. We will
show that any finite number of mesons in the discrete spectrum of the
intermediate phase become degenerate in mass at the critical
(overheated) embedding (our findings avoid the no-crossing theorem of
Von Neumann \& Wigner~\cite{Neumann:1929a} because a finite change in
the temperature leads to an infinite change of the potential). 

A similar degeneracy occurs, but only in the D3/D7 case and the polar
embedding of the D4/D6 systems (see figure~\ref{f:setup}), for the
quasi-normal frequencies when the critical embedding is approached
from the high-temperature side (undercooled). Since the masses and
quasi-normal modes are labelled by a single excitation number~$n$,
there is in principle an infinite number of ways to relate the states
in these two spectra, and the connection between mesons and
quasi-normal modes thus seems more subtle than previously
suggested~\cite{Myers:2007we,Erdmenger:2007jab}.  For the D4/D8 and
equatorial D4/D6 systems, the quasi-normal modes never degenerate, and
there is thus no smooth connection to the meson spectrum.

\section{Setup and numerical motivation}

Before we discuss our main analytic results for the behaviour of the
meson spectrum near the melting transition, let us first set the stage
and motivate our approach by some numerical results. 

\begin{figure}[t]
\begin{center}
\includegraphics[width=.5\textwidth]{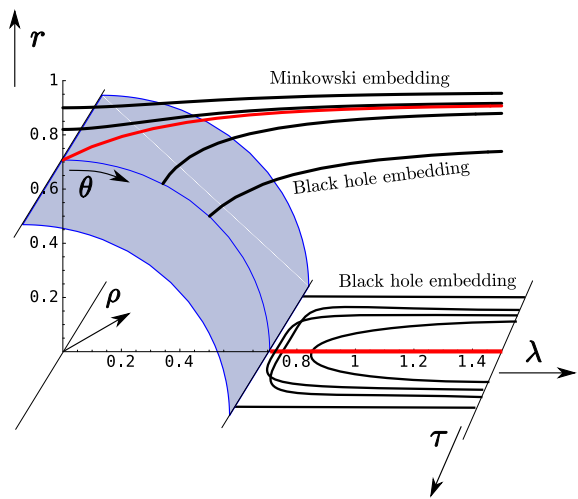}
\end{center}
\caption{The generic holographic setup for the D4/D6 system. The
  $r$--$\lambda$ plane is a conformally flat plane arising from the
  second term in~\protect\eqref{e:bgmetric}, and~$\tau\equiv x^4$ (see
  the appendix for details). The embeddings in the horizontal
  $\tau$--$\lambda$ plane will be called ``equatorial''. The
  embeddings of the D4/D8 system are similar to these equatorial ones.
  The D3/D7 system only has embeddings corresponding to the ones
  in the vertical $r$--$\lambda$ plane, which we will call
  ``polar''. Critical embeddings are displayed in red, while the
  horizon is indicated by the blue surface.\label{f:setup}}
\end{figure}

Our starting
point is the generic expression for the background metric of a
D$p$-brane. It is given by
\begin{equation}
\label{e:bgmetric}
{\rm d}s^2 = \left(\frac{u}{L}\right)^{\frac{7-p}{2}}\bigg(
\!- f_p(u) {\rm d}t^2 + \delta_{ij} {\rm d}x^i {\rm d}x^j
\bigg) + \left(\frac{u}{L}\right)^{\frac{p-7}{2}}\bigg(
\frac{{\rm d}u^2}{f_p(u)} + u^2\,{\rm d}\Omega^2_{8-p}\bigg)\,,
\end{equation}
where~$i,j=1\ldots p$\/ and~$f_p(u) = 1 -
(u_T/u)^{7-p}$. There is also a non-trivial dilaton
\begin{equation}
e^{-\phi} = \left(\frac{u}{L}\right)^{(p-7)(p-3)/4}\,.
\end{equation}
The temperature of this background is given by
\begin{equation}
T = \frac{7-p}{4\pi\,L} \left(\frac{u_T}{L}\right)^{\frac{5-p}{2}}\,.
\end{equation}
Probe branes can be embedded in various ways, depending on the value
of~$p$. In all cases which we will discuss, four of the directions of
the probe brane are aligned with~$t$ and three of the~$x^i$. For the
D4/D6 system one has a choice of two different subspaces for the
embedding of the probe.  One of these occurs again for the D3/D7
system and the other one occurs for the D4/D8 system. See
figure~\ref{f:setup} for details.

There are two types of embeddings of the probe brane, depending on
whether or not it intersects the horizon of the background. The
``Minkowski embeddings'' do not reach the horizon. The non-equatorial
embeddings of this type are described most conveniently in
the~$r$--$\lambda$ coordinate system using the~$\lambda$ coordinate on
the world-volume. The ``black hole embeddings'', on the other hand, do
reach the black hole horizon, intersecting it orthogonally. The
non-equatorial embeddings of this type are best described in a polar
coordinate system $\rho$--$\theta$, as indicated in
figure~\ref{f:setup} (see appendix~\ref{a:frames} for details on our
coordinate systems). The
embedding which lies along the horizontal axis plays a special role,
as it occurs in all models which we analyse here.

Statements concerning a possible relation between meson modes and
quasi-normal modes have all been concerned with the behaviour of
embeddings in the~$r$--$\lambda$ plane, in other words, with the D3/D7
and D4/D6 system. There is, however, also a critical embedding
separating a Minkowski from a black hole phase in the D4/D8 system,
i.e.~in the $\tau$--$\lambda$ plane. We will analyse this system
separately in section~\ref{s:D4D8}.

Let us now turn to the vector fluctuations on the probe branes. The
equations governing them follow from an expansion of the induced
action in powers of the abelian field strength. At lowest order one
has
\begin{equation}
S_{\text{vector}} \propto \int\!{\rm d}^4x{\rm d}\lambda{\rm d}\Omega_{6-p}\,
e^{-\phi} \sqrt{-\hat{g}}\, F_{MN} F_{PQ} \hat{g}^{MP} \hat{g}^{NQ}\,,
\end{equation}
(and similar for the~$\rho$--$\theta$ and~$\tau$--$\lambda$ system)
where~$\hat{g}$ denotes the induced metric.  The spatial components of
the fields which describe the massive vector mesons arise from the
expansion
\begin{equation}
F_{0 a} = \sum_n \partial_0 B_a^{(n)}\,\psi_{(n)}(\lambda)\,,\qquad
F_{\lambda a} = \sum_n B^{(n)}_a \,\partial_\lambda\psi_{(n)}(\lambda)\,.
\end{equation}
We will focus solely on thermal pole masses of the vector mesons,
which are defined by $\partial_0^2 B_a^{(n)} = - m_{(n)}^2 B_a^{(n)}$.
The equation of motion for the modes~$\psi_{(n)}$ which then follows
reads
\begin{equation}
\label{e:fluceq}
\partial_{\lambda}
\left[ e^{-\phi}\sqrt{-\hat{g}} \hat{g}^{aa} \hat{g}^{\lambda\lambda}
  \partial_{\lambda}{\psi}_{(n)}\right]
+ m_{(n)}^2 e^{-\phi}\sqrt{-\hat{g}} \hat{g}^{aa} \hat{g}^{00} \psi_{(n)} = 0\,.
\end{equation}
These equations can be solved in a series expansion near the tip of
the brane, and then extended numerically using standard shooting
methods. In the Minkowski phase, the discrete mass spectrum arises
after imposing that the fluctuations are regular in the IR and
normalisable in the UV. In the black hole phase, the discrete
quasinormal spectrum comes from incoming boundary conditions in the IR
and normalisability in the UV.

In order to get a better insight into the structure of the spectrum,
it is useful to perform a coordinate transformation and a wave
function rescaling which brings~\eqref{e:fluceq} in Schr\"odinger
form. Following the notation of~\cite{Schreiber:2004ie}, we write the
equation in the general form
\begin{equation}
\frac{1}{\Gamma(\lambda)}\partial_{\lambda} \left[
  \frac{\Gamma(\lambda)}{\Sigma^2(\lambda)} \partial_\lambda\psi_{(n)} \right]
+ m^2_n\,\psi_{(n)} = 0\,.
\end{equation}
(The functions~$\Gamma$ and $\Sigma$ of course implicitly depend on
the probe brane embedding function~$r(\lambda)$).  By introducing a new
coordinate and a rescaled wave function according to
\begin{equation}
\frac{{\rm d}\lambda}{{\rm d}\sigma} = \frac{1}{\Sigma}\,,\qquad
\tilde\psi_{(n)} = \Xi\, \psi_{(n)}\,,\qquad
\Xi = \sqrt{\frac{\Gamma}{\Sigma}}\,,
\end{equation}
the problem is transformed into a Schr\"odinger type problem,
\begin{equation}
- \partial^2_{\sigma}\tilde{\psi}_{(n)}
 + V\tilde{\psi}_{(n)} = E_n \tilde{\psi}_{(n)}\,,\quad\text{with}\qquad
V = \frac{\partial_{\sigma}^2\Xi}{\Xi}\,,\quad E_n = m_{(n)}^2\,.
\label{eq schro1}
\end{equation}
We will see how to approximate the potential~$V$ analytically in the
following sections. However, to get an idea of what we are aiming for,
let us first give the results of a numerical analysis, valid for the
embeddings in the~$r$--$\lambda$ plane.

\begin{figure}[t]
\includegraphics[width=.32\textwidth]{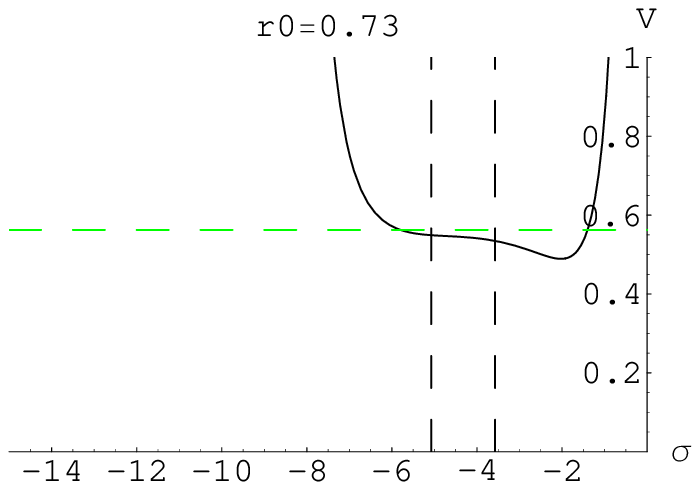}
\includegraphics[width=.32\textwidth]{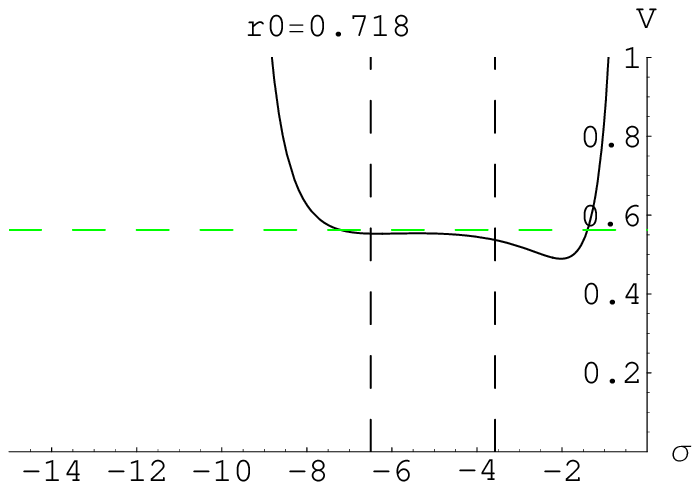}
\includegraphics[width=.32\textwidth]{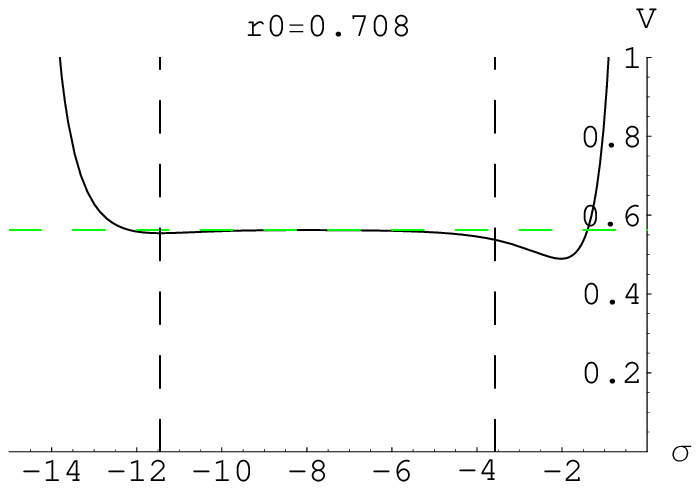}
\caption{The Schr\"odinger potential for Minkowski embeddings in the
  D3/D7 system, close to criticality, for various values~$r_0$ of the
  intersection with the vertical axis. The green dashed line indicates
  the asymptotic potential in the box, given
  by~\protect\eqref{e:Vcrit}; the vertical dashed lines mark the edges
  of the plateau (which will be defined in section~\protect\ref{s:SchrStruct}).\label{f:V_Mink}}
\end{figure}

\begin{figure}[t]
\begin{center}
\vspace{3ex}
\includegraphics[width=.32\textwidth]{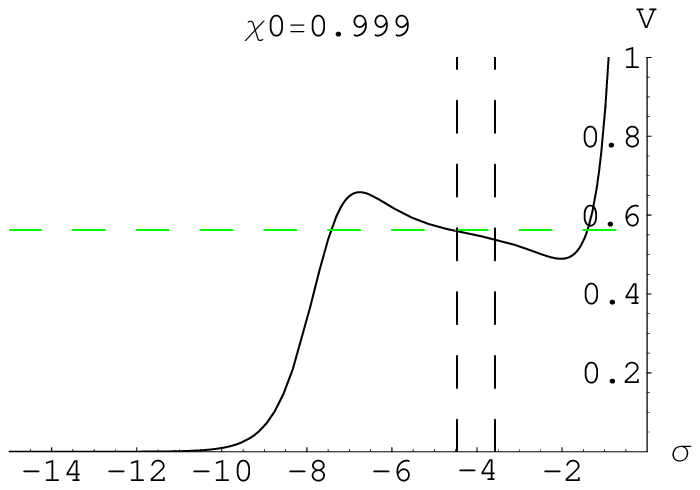}
\includegraphics[width=.32\textwidth]{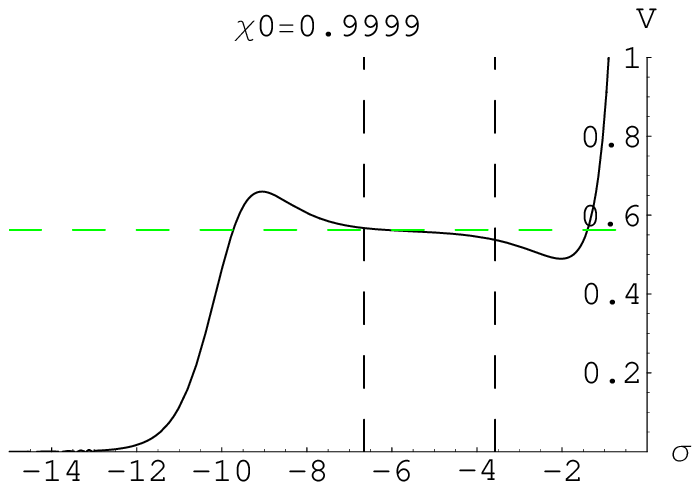}
\includegraphics[width=.32\textwidth]{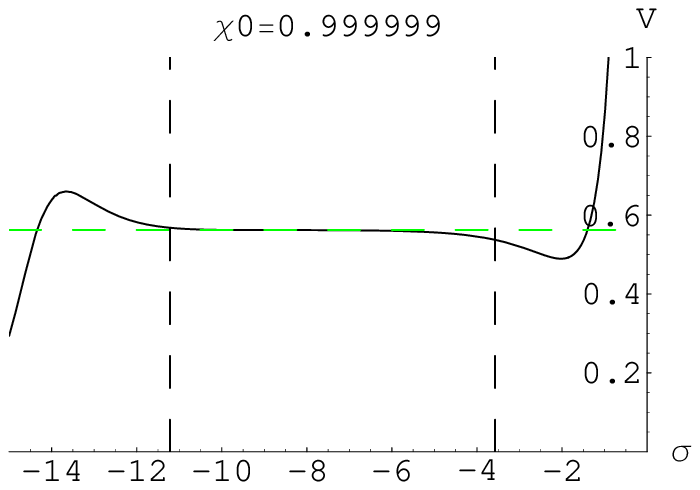}
\end{center}
\caption{The Schr\"odinger potential for black hole embeddings in the
  D3/D7 system, close to criticality, for various values of the
  angle~$\chi_0$ of intersection with the horizon. Green and black
  dashed lines have the same meaning as
  in~\protect\ref{f:V_Mink}.\label{f:V_bh}}
\end{figure}

The resulting potential for the Minkowski embedding is displayed in
figure~\ref{f:V_Mink}, where~$r_0$ is the position at which the probe
brane intersects the vertical axis of
figure~\ref{f:setup}. The~$\sigma$ coordinate is bounded on both
sides. As the critical embedding is approached, the potential becomes
more and more rectangular, while the box size (i.e.~the range
of~$\sigma$) grows, until it eventually becomes unbounded.

The result for the black hole embedding is shown in
figure~\ref{f:V_bh}, where~$\chi_0 = \cos\theta_0$ is related to the
value of the angular coordinate at which the brane intersects the
horizon, as shown in figure~\ref{f:setup}. Now the~$\sigma$
coordinate is bounded only in the UV direction, and the potential 
develops a clear step-shape. As the critical embedding is approached,
the step becomes infinitely long.

These figures suggest that it should be possible to find a simple
description of the potential very close to the critical embedding. We
will do that in the next section. A somewhat similar story holds true
for the D4/D8 embeddings, which we will analyse in more detail in
section~\ref{s:D4D8}.

\section{The D3/D7 system}
\subsection{Zero-temperature potentials}

Before giving our analytic results at finite temperature, it is
instructive to recall the structure of the Schr\"odinger potential at
zero temperature. Although we mainly focus on the D3/D7 system
($p=3$), we have kept part of the discussion general so the D4/D6
system can be obtained (by setting $p=4$).  At zero temperature, there
are two types of embeddings. One is the conformal embedding of the
D7-brane (when the constituent quark mass $m_q$ is zero), and the
other one is the non-conformal embedding, when $m_q \neq 0$.  In the
former case the induced metric on the worldvolume of the brane is
exactly that of $\text{AdS}_5 \times S^3$, while in the latter case
the induced metric approaches this only in the UV, i.e.~near the AdS
boundary. The embedding is trivial in these cases, given by~$r=m_q$,
which leads to the vector fluctuation
equation~\cite{Kruczenski:2003be}\footnote{The curvature
  radius~$L$ has been set to one, hence the~$m_q$ variable has
  dimension length. The coordinates used are \mbox{$r=u\cos\theta$} and~$\lambda=u
  \sin\theta$ as in~\cite{Kruczenski:2003be}, but different from the
  coordinates of appendix~\ref{a:frames}.}
\begin{equation}
\frac{1}{\lambda^3} \partial_\lambda \Big( \lambda^3
\partial_\lambda\psi_{(n)}\Big) + \frac{m_{(n)}^2}{(\lambda^2+m_q^2)^2} \psi_{(n)}  = 0\,.
\end{equation}
The Schr\"odinger potentials for the~$m_q=0$ and $m_q>0$ are easily
computed to be, 
\begin{equation}
\label{VT0}
\begin{aligned}
V_{m_q=0}(\sigma) & =  \frac{3}{4\,\sigma^2}\,, & \lambda &= -\sigma^{-1}\,,
\\[1ex]
V_{m_q\not=0}(\sigma) & = m_q^2 \left[ \frac{1}{2} + \frac{3}{4}\Big(
  \tan^2(m_q \sigma) +
\frac{1}{\tan^2(m_q \sigma)}\Big)\right]\,, & \lambda &= m_q \tan(m_q\sigma) \,.
\end{aligned}
\end{equation}
In the conformal case the potential goes to a constant (zero) in the
IR, while it diverges in the UV.  In the non-conformal case, the
potential diverges both in the UV and IR. The box shape of the
potential $V_{m_q\not=0}$ obviously leads to a discrete spectrum, in
contrast to the continuum spectrum of the conformal case. If one
replaces~$V_{m_q\not=0}$ with an infinitely high box of
size~$\pi/(2m_q)$, the spectrum asymptotically becomes $m_{(n)}^2 = 4
m_q^2\, n^2$, in agreement with the large-$n$ limit of the exact
spectrum found in~\cite{Kruczenski:2003be}. Hence, we see that while
in the UV both branes are the same (i.e.~AdS), with a diverging
potential, the main difference between the two comes from the IR,
i.e.~from near the Poincar\'e horizon.

Let us now turn to the $T\neq 0$ case. In this case the analytic form
of the embeddings of the branes is unknown, and so one would perhaps
expect the analysis to be more complicated than above. However, the
key element is that, for branes close to the critical embedding, only
the near horizon region of the background is relevant, a region which
is quite different from the $T=0$ background discussed above. Brane
embeddings near the horizon were analysed
by~\cite{Frolov:1998td,Mateos:2006nu}. As we will show below, when we
get closer and closer to the critical embedding, the region in
Schr\"odinger coordinate which corresponds to the near-horizon region
becomes larger and larger, growing without bound. In the strict limit,
we end up with an extremely simple, flat potential of infinite length,
only corrected by finite-size effects whose relative effect scales to
zero. The simple potential then allows us to make an exact statement
about the spectrum.

\subsection{Brane embeddings in the Rindler approximation}
\label{sec: Rindler}

Having illustrated the behaviour of the Schr\"odinger potential with
some numerical results and having sketched our approach, let us now
provide an analytic argument which will lead to an exact potential in
the critical limit. We first need to summarise the brane embeddings
close to the horizon. As in~\cite{Mateos:2006nu}, we write the metric of
the $8-p$ sphere as
\begin{equation}
{\rm d}\Omega_{8-p}^2 = {\rm d}\theta^2 + \sin^2 \theta {\rm d}\Omega_{6-p}^2 +
\cos^2\theta {\rm d}\varphi^2\,.
\end{equation}
We will consider the probe embedding extended in three~$x$-directions,
wrapping the~$6-p$-sphere at fixed $\varphi$ and having non-trivial
$\theta(r)$.  We now want to zoom in a region near the horizon, for a
brane which is near its critical embedding. This means a region close
to $\theta = 0$, $r=r_0$. Let us choose a parametrisation
\begin{equation}
u = u_T + \pi T z^2\,,\qquad
\theta = \frac{y}{L} \left(\frac{L}{u_T}\right)^\frac{p-3}{4}\,,\qquad
\tilde x = \left(\frac{u_T}{L}\right)^\frac{7-p}{4} x\,.
\end{equation}
Then, considering the limit where
\begin{equation}
\frac{u-u_T}{u_T} = \frac{\pi T z^2}{u_T}\ll 1 \,\,,\qquad 
\theta = \frac{y}{L} \left(\frac{L}{u_T}\right)^\frac{p-3}{4} \ll 1\,,
\end{equation}
we find Rindler space
\footnote{The relation with the~$\rho,\chi$ variables of
  figure~\ref{f:setup} is, for the D3/D7 system, given by $\rho - 1 = z/L$,
$\chi = 1 - \frac12 y^2/L^2$, valid near $\rho \approx 1$, $\chi
  \approx 1$; see appendix~\ref{a:frames} for details.\label{fn:rhochi_zy}}
\begin{equation}
{\rm d}s^2 = -(2\pi T)^2 z^2 {\rm d}t^2 + {\rm d}z^2 + {\rm d}y^2
+ y^2 {\rm d}\Omega_{6-p}^2 + {\rm d}\tilde x_3^2 + \dots\,,
\end{equation}
while the dilaton approaches a constant.  Notice the horizon is now at
$z=0$.  The Rindler approximation ceases to be valid at distances set
by
\begin{equation}
R_{\text{Rindler}} \equiv  \sqrt{\frac{u_T}{\pi T}}  = \begin{cases} L
  \, \quad & p=3;
  \\[1ex]
\frac{2}{\sqrt{3}} \left(u_T L^3\right)^{1/4} \, \quad & p=4 \, .
\end{cases}
\end{equation}

The equation describing the probe embedding $y(z)$ in this region is
\begin{equation}
z y \ddot y + \big(y \dot y - (6-p) z\big) \big(1+ \dot y^2\big) = 0\,.
\label{RindlerEmbedding}
\end{equation}
An important aspect of~\eqref{RindlerEmbedding} is its scaling
symmetry, $z \to \lambda z$, $y \to \lambda y$. This means that
changing $z_0$ or $y_0$ does not really influence the shape of the
embedding but just rescales it. 

The obvious solution of this equation gives the critical embedding, 
\begin{equation}
\label{yc}
y_{\text{crit}} = \sqrt{6-p}\,z \, .
\end{equation}
If the brane does not touch the horizon, there is a family of smooth
Minkowski embeddings parametrised by $z_0$, the distance of the tip
of the brane to the horizon. The solution near $z=z_0$ reads
\begin{equation}
\label{nBH}
y_{\text{Mink}} = z_0 \left[
 \sqrt{2(7-p)} \sqrt{\frac{z-z_0}{z_0}} + {\cal O}\left(\left(\frac{z-z_0}{z_0}\right)^\frac32\right)\right]\,.
\end{equation}
For $z \gg z_0$ one can find the asymptotic
solution~\cite{Frolov:1998td} which shows that the embedding
approaches the critical one~$y_{\text{crit}}$ up to corrections which
scale as a negative power of~$z$.  Taking the limit $z_0 \rightarrow
0$ corresponds to approaching the critical embedding.\footnote{However,
  note that the limit is subtle.  Since the expansion parameter is
  $(z-z_0)/z_0$ one cannot first take $z_0 \rightarrow 0$, but rather
  one has to simultaneously scale $z\rightarrow 0$. This is similar to
  the situation we had for the $T=0$ embedding and the limit
  $m_q\rightarrow 0$.}  If the brane intersects the horizon, there is
a family of black hole embeddings parametrised by $y_0$, which near
$z=0$ reads
\begin{equation}
\label{BH}
y_{\text{BH}} = y_0 \left[ 1 + \frac{6-p}{4}\frac{z^2}{y_0^2} + {\cal O}\left(
\left(\frac{z}{y_0} \right)^3 \right)\right]\,.
\end{equation}
Again, the parameter $y_0$ measures the deviation from the critical
embedding, and taking the limit $y_0 \rightarrow 0$ corresponds to the
limit in which the critical embedding is reached.  In the other
regime~$z\gg z_0$ the embedding again approaches the critical one~\cite{Frolov:1998td}.

\subsection{Structure of the Schr\"odinger potentials}
\label{s:SchrStruct}

Let us now analyse the Schr\"odinger potentials for the spectrum of
vector meson fluctuations. The dominant part of the potential will
turn out to be located in the Rindler region, so we will first focus
on the embeddings described in the previous
subsection.\footnote{Several other properties of holographic QCD are
  also fully determined by the Rindler region; see
  e.g.~\cite{Peeters:2007ti}.}  We will focus on a simple subset of
the possible excitations, namely those described (in the $A_0 = 0$
gauge) by $A_i = e^{-i \omega t} \psi (z)$. In Rindler coordinates,
the equation for the fluctuation is
\begin{equation}
\partial_z \left(y^{6-p} z (1 + \dot y^2)^{-\frac12} \partial_z
\psi\right) + \frac{y^{6-p}}{z} (1 + \dot y^2)^{\frac12}
 {\tilde \omega}^2 \psi =0\,,
\label{eqpsi}
\end{equation}
where $\tilde \omega \equiv \frac{\omega}{2 \pi T}$.
Using~\eqref{RindlerEmbedding} this equation can be written as
\begin{equation}
\partial_z^2 \psi + \frac{1 + \dot y^2}{z} \partial_z \psi
+ \frac{1 + \dot y^2}{z^2} \tilde \omega^2 \psi = 0\,.
\end{equation}
The IR boundary conditions to be imposed on the wave function
$\psi$ are regularity in the Minkowski case and incoming boundary
conditions in the black hole case,
\begin{equation}
\psi_{\text{Mink}} = 1 + {\cal O} (z-z_0)\,,\qquad
\psi_{\text{bh}} = z^{-i \tilde \omega} (1 + {\cal O} (z))\,.
\end{equation}
It is useful to recast equation~\eqref{eqpsi} in Schr\"odinger
form, as we did for the full problem in the previous section. In terms
of the notation used in~\cite{Schreiber:2004ie} we now have
\begin{equation}
\Gamma = \frac{y^{6-p}}{z}(1+\dot y^2)^\frac12\,,\qquad
\Sigma = \frac{(1+\dot y^2)^\frac12}{z}\,,\qquad
\Xi = y^\frac{6-p}{2}\,,\qquad
\tilde{\psi} = \Xi \psi\,.
\label{definethings}
\end{equation}
The new radial variable $\sigma$ is defined through ${\rm
  d}\sigma/{\rm d}z = \Sigma$, as before. With these redefinitions,
equation~\eqref{eqpsi}, describing fluctuations in the IR,
can be rewritten as
\begin{equation}
\label{schroeq}
-\partial_{\sigma}^2 \tilde{\psi} + V_{\text{IR}} \tilde{\psi} = \tilde \omega^2
\tilde{\psi}\,,\qquad
\text{with}\qquad
V_{\text{IR}} = \frac{\partial_\sigma^2 \Xi}{\Xi} = \frac{(6-p)\ z^2}{2 y^2}\
\frac{6-p+((6-p)/2 -1) (\partial_z y)^2}{1+ (\partial_z y)^2}\,.
\end{equation}
Using the explicit solutions for brane embeddings, described in the
previous subsection, we can compute the potentials. They are plotted
in figure~\ref{f:estimates} for the~$p=3$ case.

\begin{figure}[t]
\begin{center}
\includegraphics[width=.8\textwidth]{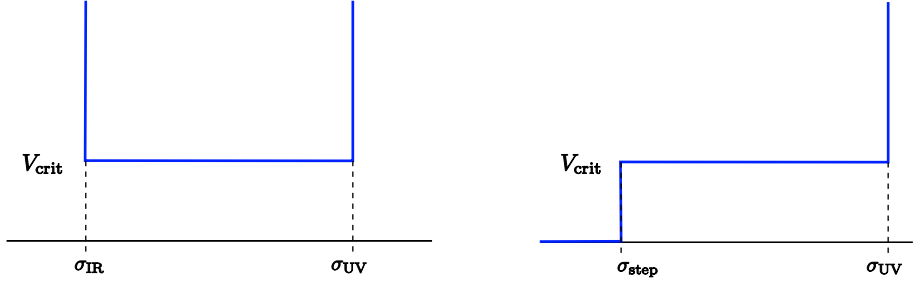}
\end{center}
\caption{The approximations used for the potentials close to the
  critical embedding, to be compared with the exact numerical results
  in figure~\protect\ref{f:V_Mink} and~\protect\ref{f:V_bh}.\label{f:approxV}}
\end{figure}

Let us first discuss the potential for the critical embedding~\eqref{yc}. It
turns out to be constant everywhere in the Rindler region, with a
value given by
\begin{equation}
\label{e:Vcrit}
V_{\text{crit}}^{\text{IR}} = \frac{(6-p)^2}{4(7-p)}\,.
\end{equation}
The $\sigma$ coordinate is related to the~$z$ coordinate by $\sigma =
\sqrt{7-p} \log z$, so the point $z=0$ is mapped to $\sigma=-\infty$
while the edge of the Rindler region in the UV is mapped
to~$\sigma=+\infty$. The potential for the critical embedding is thus
a constant on $\sigma=\langle -\infty, +\infty\rangle$.\footnote{If
  one repeats the computation for a scalar excitation $y =
  y_{\text{cl}} (z) + \delta y (t, z)$, one gets for the critical
  embedding $V_{\text{crit}} = \frac{p^2 - 8 p + 8}{4 (7-p)^2}$ which
  is negative for the cases of interest $p=3,4$, signalling
  unambiguously the appearance of scalar tachyons in this channel near
  the critical embedding. These tachyonic modes were first explicitly
  discussed in \cite{Mateos:2007vn}.}

We have already seen numerical evidence for the formation of this
plateau both from the Minkowski side as well as from the black hole
side (figure~\ref{f:V_Mink} and~\ref{f:V_bh}). In the Minkowski case,
the plateau arises from an approximately square well, while in the
black hole case it arises from a step-like potential (note that the
step is a generic feature of black hole
embeddings~\cite{Horowitz:1999jd}). We thus want to approximate these
potentials as in figure~\ref{f:approxV}.\footnote{For Minkowski
  embeddings in the D4/D6 case, the Schr\"odinger potential does not
  diverge at $z=z_0$ but goes to a finite constant. However,
  regularity of the fluctuation requires $\tilde \psi$ to vanish at
  the finite value $\sigma_{\text{IR}}=\sigma (z=z_0)$. Thus, a box
  potential is also a good approximation in this case.} Using the
results of the present section, we can now make explicit how the
\emph{lengths} of the long flat segments of these potentials depend on
the \emph{distance} to the critical embedding.

\begin{figure}[t]
\begin{center}
\includegraphics[width=.5\textwidth]{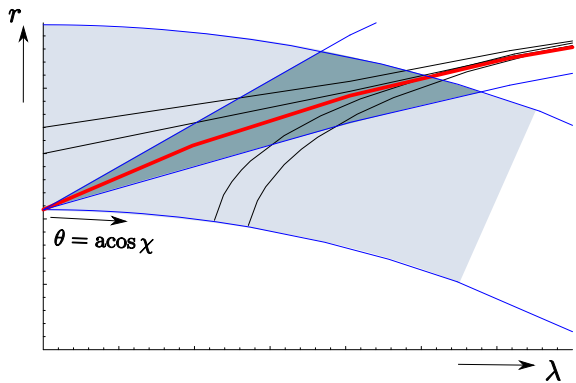}
\end{center}
\caption{Schematic indication of the three regions defined in the
  text. The light shaded area is the Rindler region. For a fixed
  embedding, the segment before entering the dark shaded area is the
  ``deep IR'', the segment after exiting is the ``UV'' while the
  segment inside the dark shaded area is the ``intermediate IR''. The
  latter grows, in the Schr\"odinger coordinate, 
  as~$-2\log(r_0 - 1/\sqrt{2})$ or $-\log(1-\chi_0)$.\label{f:region3}}
\end{figure}

\begin{figure}[t]
\begin{center}
\includegraphics[width=.43\textwidth]{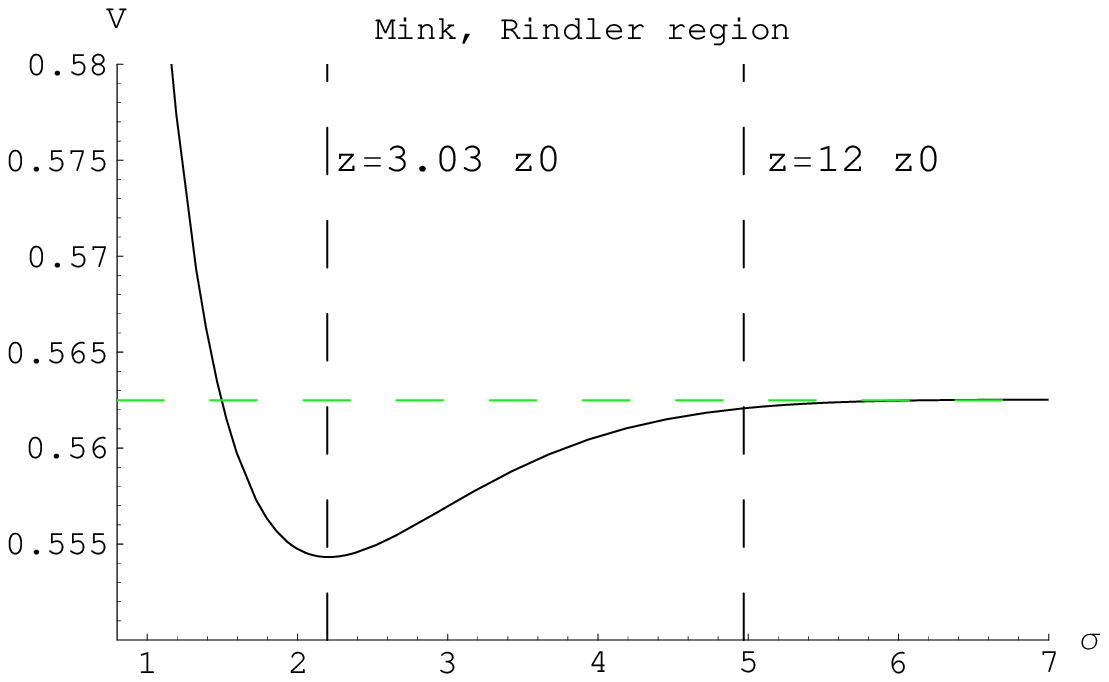}\qquad
\includegraphics[width=.43\textwidth]{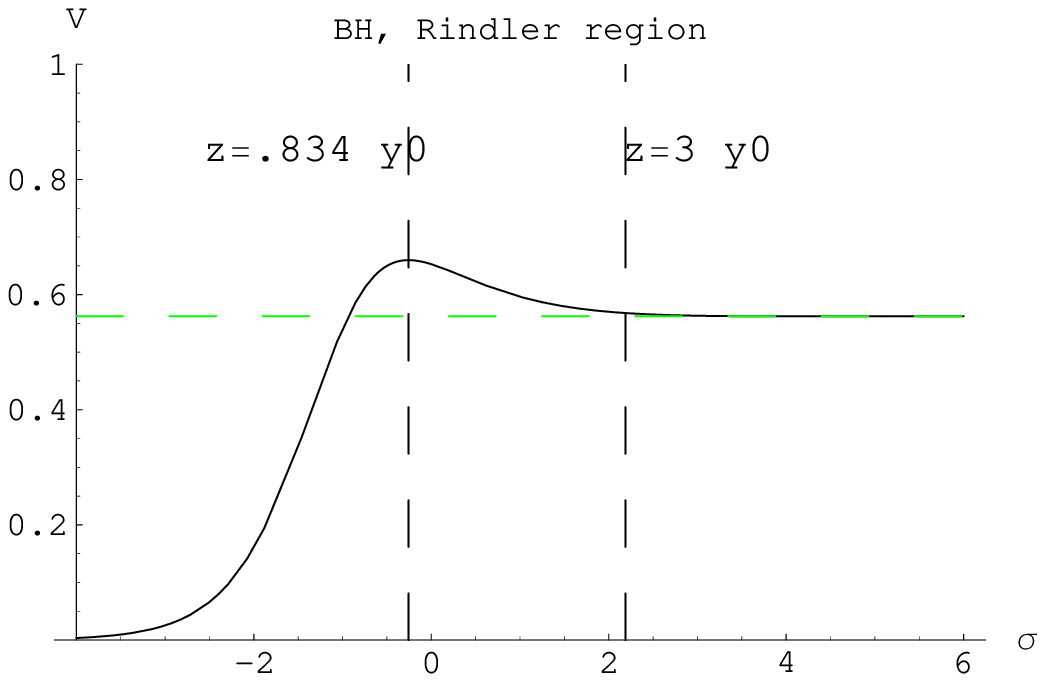}
\end{center}
\caption{Schr\"odinger potentials for the D3/D7 vector excitation in Rindler
  space, as well as an estimate of the left-hand side of the plateau,
  for the Minkowski and black hole embeddings
  respectively.\label{f:estimates}}
\end{figure}

For all embeddings, three regions can be distinguished. First, there
is a ``deep IR region'', where the near-critical Minkowski and
near-critical black hole embeddings still deviate from the critical
embedding. Somewhat further out, but still in the region where the
Rindler approximation makes sense, there is an ``intermediate IR
region'' or ``plateau''. In this region the embeddings are
well-approximated by the critical embedding.  Finally, for~$z\approx
R_{\text{Rindler}}$, we leave the regime of validity of the Rindler
approximation, and we have no analytic control anymore over the shape
of the embeddings in this ``UV region''. The three regions are shown
in figure~\ref{f:region3}.

The potential in the intermediate IR region is essentially constant,
because the brane is there almost the same as the critical one. The
key observation is that, as we argue below, the size of this
intermediate region in the Schr\"odinger coordinate diverges
logarithmically as the critical embedding is approached. The other two
regions (``deep IR'' and ``UV region''), on the other hand, have a
size which is independent of~$z_0$ or $y_0$. Hence, as observed
numerically, when we approach the critical embedding (either from the
black hole or from the Minkowski side) the Schr\"odinger potential
becomes constant, due to the fact that the size of the intermediate IR
region in the Schr\"odinger coordinate diverges.

Let us now compute the lengths of the various regions.  First, in order to
show that the \emph{deep IR region} has a size which is independent
of~$z_0$ or~$y_0$, we make use of the scaling symmetry
of~\eqref{RindlerEmbedding}.  Suppose that we have a Minkowski
solution
\begin{align}
  y &= f_1(z)\,,&&\hspace{-6em}\text{which satisfies \quad
    $y(\bar{z}_0)=0$,} 
\intertext{i.e.~$\bar{z}_0$ is the point where the brane is
  closest to the horizon. Using scaling symmetry, there is
  then another solution}
y &= f_2(z) \equiv f_1\left(z
\frac{\bar{z}_0}{z_0}\right) \frac{z_0}{\bar{z}_0}\,,
&&\hspace{-6em}\text{which satisfies $y(z_0)=0$.}
\end{align}
Next, let us introduce a threshold~$\Delta$, which is a fixed and
small positive number. When the distance between two branes is smaller
than~$\Delta$, we will consider them to be coincident.  Assume now
that we have found a point \mbox{$z=\bar{z}_{\text{left}}$} for which
the distance between the first embedding and the critical one is equal
to this~$\Delta$,
\begin{equation}
f_1(\bar{z}_{\text{left}}) - \sqrt{6-p}\, \bar{z}_{\text{left}} = \Delta\,.
\end{equation}
This tells us immediately that, at a point~$z_{\text{left}} = 
(\bar{z}_{\text{left}}/\bar{z}_0)\,z_0$, the distance of the second embedding to
  the critical one satisfies
\begin{equation}
f_2(z_{\text{left}}) - f_{\text{crit}}(z_{\text{left}}) = 
f_1\left( z_0 \frac{\bar{z}_{\text{left}}}{\bar{z}_0}\frac{\bar{z}_0}{z_0}\right)
\frac{z_0}{\bar{z_0}} - \sqrt{6-p}\, z_0 \frac{\bar{z}_{\text{left}}}{\bar{z}_0}
= \frac{z_0}{\bar{z}_0} \Delta < \Delta\qquad\text{if $z_0<\bar{z}_0$.}
\end{equation}
In other words: if we define the end of the deep
IR region to scale like linearly in $z_0$, and if we fix the
proportionality constant for some value of $z_0$, then the
approximation is better for all embeddings which are closer to
critical. We can thus define the deep IR region to be
$z_0 < z < \Lambda z_0$ for some~$\Lambda$ which is independent of~$z_0$.

The length of the deep IR region in the Schr\"odinger
coordinate~$\sigma$ now comes out to be independent of~$z_0$,
because~$\partial_z y$ is a function of $z/z_0$ only (this is a
consequence of the scaling symmetry). Therefore, using the value
of~$\Sigma$ in~\eqref{definethings}, we find
\begin{equation}
\Delta\sigma_{\text{deep IR}}^{\text{Mink}} = \int_{z_0}^{\Lambda z_0}
\frac{\big(1 + Y^2(z/z_0)\big)^{1/2}}{z} {\rm d}z = \int_1^{\Lambda}
\frac{\big(1+ Y^2(q)\big)^{1/2}}{q} {\rm d}q\,,
\end{equation}
where $Y(z/z_0)\equiv\partial_z y$.  This expression is independent
of~$z_0$.  The length of the ``intermediate IR'' region,~$\Lambda z_0
< z < \epsilon R_{\text{Rindler}}$ for some~$\epsilon < 1$, can be
computed using the critical embedding.

\begin{figure}[t]
\begin{center}
\hspace{-0.7cm}\includegraphics[height=5cm]{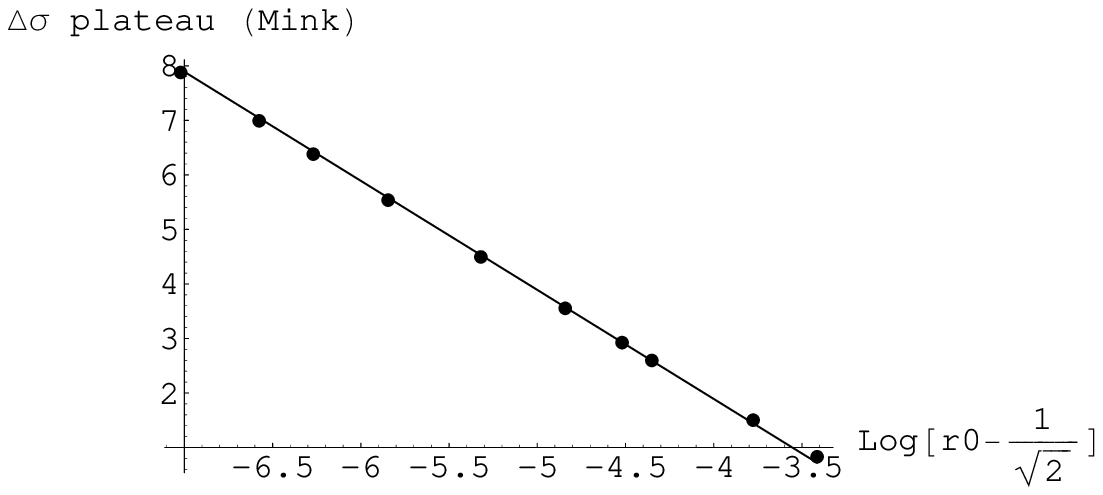}\qquad
\raisebox{3.8ex}{\includegraphics[height=3.78cm]{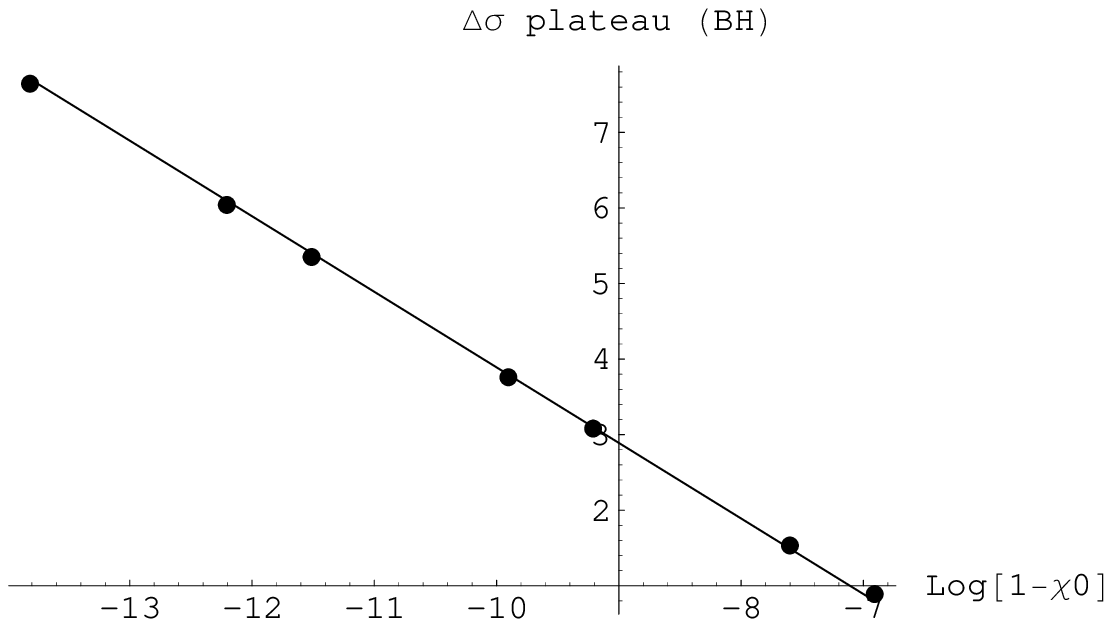}}
\end{center}
\caption{Numerical values (for the D3/D7 system) for the width of the
  plateau in the Schr\"odinger~$\sigma$ coordinate, for the Minkowski
  and black hole embeddings, together with their analytic
  approximations~\protect\eqref{e:widthestimateMink}
  and~\protect\eqref{e:widthestimate}.\label{f:plateauwidth}}
\end{figure}

We now concentrate on the D3/D7 case, for which the edges of the
plateau are given by~$\Lambda \approx 3$ and~$\epsilon \approx
1/5$. One finds
\begin{equation}
\label{e:widthestimateMink}
\Delta \sigma_{\text{plateau}}^{\text{Mink}} \approx \int_{\Lambda z_0}^{\epsilon L} \frac{2}{z}{\rm
  d}z = 2 \log \frac{\epsilon L}{\Lambda z_0} \approx 
-2\log(r_0-1/\sqrt{2}) - 6.11\,.
\end{equation}
The numerical factors~$\epsilon$ and~$\Lambda$ are of course somewhat
arbitrary, but the scaling with~$r_0$ is fixed by the critical
embedding. This scaling is reproduced well by the numerical data, see
figure~\ref{f:plateauwidth}a (the black dashed vertical lines in
figure~\ref{f:V_Mink} indicate the edges of the plateau).  Finally,
there is a UV region which is out of reach of the Rindler
approximation. However, this UV region is similar to the one for
zero-temperature embeddings, i.e.~has finite size in the Schr\"odinger
coordinate, independent of~$r_0$.

A similar logic applies to the black hole embeddings. The agreement
with the critical embedding (or equivalently, the start of the
plateau) again occurs at around~$z\approx 3y_0$, and the Rindler
approximation breaks down at~$z\approx L/5$. We then find that the
plateau region scales as
\begin{equation}
\label{e:widthestimate}
\Delta \sigma_{\text{plateau}}^{\text{BH}} \approx \int_{\Lambda y_0}^{\epsilon L} \frac{2}{z}{\rm
  d}z = 2 \log \frac{\epsilon L}{\Lambda y_0}  \approx 
-\log(1-\chi_0) - 6.11\,,
\end{equation}
Again, this scaling with~$\chi_0$ agrees well with the size of the
plateau as measured from the numerical data, see
figure~\ref{f:plateauwidth}b (the black dashed vertical lines in
figure~\ref{f:V_bh} indicate the edges of the plateau). There is also
again a UV region, which looks like the UV part of the potential at
zero-temperature, which has finite size in the Schr\"odinger
coordinate.  

We thus see that in both cases, we can determine the size of the
plateau as a function of the distance to the critical embedding. The
plateau becomes of infinite length, and all other IR and UV features
of the potentials get infinitely pushed away. They remain of finite
size, however, and their contributions to the spectrum scale to zero
in the critical limit (some caricature potentials in which this can be
seen explicitly are discussed in appendix~\ref{appendixB}). Therefore,
we conclude that the mass spectrum can be approximated by using a
square well potential on the Minkowski side, and a square step when
one approaches from the black hole side (see figure~\ref{f:approxV}),
where the plateau sizes scale as in~\eqref{e:widthestimateMink}
and~\eqref{e:widthestimate}.

\subsection{Meson and quasi-normal modes}

Having obtained simple potentials for the two types of brane
embeddings, let us now look at the spectrum.  For the Minkowski
embedding, the masses are simply given by the solution of the
Schr\"odinger equation in a box of height~$V_{\text{crit}}$. This yields
\begin{equation}
\sqrt{\tilde \omega_n^2 - V_{\text{crit}}} \approx \frac {n
  \pi}{\sigma_{\text{UV}} - \sigma_{\text{IR}}}\,,\qquad n \in \mathbb{N}\,.
\end{equation}
However, the box size~$\sigma_{\text{UV}}- \sigma_{\text{IR}}$ depends
on~$z_0$, and diverges in the critical limit. Thus, all meson masses
tend to~$\sqrt{V_{\text{crit}}}$, as alluded to in the
introduction. Verifying this numerically, on the other hand, is
slightly subtle because the masses approach the limit only slowly. 

For the black hole embedding, we need to consider a potential with two
regions.  In the first region, with range $-\infty < \sigma <
\sigma_{\text{step}}$, the potential vanishes. The solution of the
Schr\"odinger equation which has incoming boundary conditions
at~$\sigma=-\infty$ is $\tilde{\psi}_{(1)} = e^{-i \tilde \omega
  \sigma}$.  To simplify notation, we will
set~$\sigma_{\text{UV}}=0$. In the second region, with range
$\sigma_{\text{step}} < \sigma < 0$, the potential has the value given
in~\eqref{e:Vcrit}. The solution now reads \mbox{$\tilde{\psi}_{(2)} = A
\sin(\sqrt{\tilde \omega^2 - V_{\text{crit}}} \sigma)$} where $A$ is a constant to be determined
and we have imposed $\tilde{\psi}(\sigma=0)=0$ since there is an infinite
wall.  By matching the values of $\tilde{\psi}$ and its derivative at
$\sigma=\sigma_{\text{step}}$, we find a single equation for the
values of the quasi-normal frequencies,
\begin{equation}
i \tilde \omega \sin (\sqrt{\tilde \omega^2 - V_{\text{crit}}}\ \sigma_{\text{step}})
+ \sqrt{\tilde \omega^2 - V_{\text{crit}}} 
\cos (\sqrt{\tilde \omega^2 - V_{\text{crit}}}\ \sigma_{\text{step}}) =0\,.
\label{eqforstep}
\end{equation}
Given $V_{\text{crit}}$ and $\sigma_{\text{step}}$, one can find numerically the 
complex values of $\tilde \omega$ that solve this equation.
However, let us look at a particularly interesting limit,
\begin{equation}
|\sigma_{\text{step}}| \sqrt{V_{\text{crit}}} \gg 1\,.
\end{equation}
Since in our problem $V_{\text{crit}}$ is fixed to be the constant
in~\eqref{e:Vcrit}, this is satisfied for large (negative)
$\sigma_{\text{step}}$. Remembering that 
$|\sigma_{\text{step}}|$ grows unbounded when the flavour brane
approaches the critical embedding ($y_0 \to 0$), this is in fact the
limit in which we are interested. But now we can obtain $\tilde
\omega$ as an expansion in $(\sigma_{\text{step}} \sqrt{V_{\text{crit}}})^{-1}$,
\begin{equation}
\label{e:rindlerQNM}
\tilde \omega_n = \sqrt{V_{\text{crit}}} \left(
\pm 1 \pm \frac {n^2 \pi^2}{2 \sigma_{\text{step}}^2 V_{\text{crit}}} + 
i F \frac {n^2 \pi^2}{\sigma_{\text{step}}^3 V_{\text{crit}}^\frac32}+ 
{\cal O} \left( (\sigma_{\text{step}}\sqrt{V_{\text{crit}}})^{-4} \right)\right)
\,,\qquad n\in{\mathbb N}\,.
\end{equation}
Here~$F$ is the first coefficient in this expansion which depends on
the finite-size details of the potential; for the step potential of
figure~\ref{f:approxV} we have~$F=1$ (for details see appendix~\ref{appendixB}).

\begin{figure}[t]
\begin{center}
\includegraphics[width=.5\textwidth]{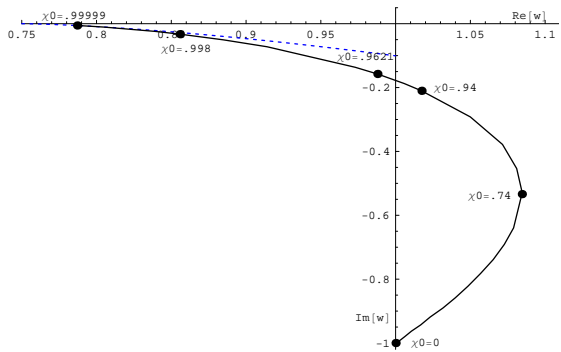}
\end{center}
\caption{Behaviour of the first quasi-normal mode of the D3/D7 system
  with a black hole embedding, as a function of the distance~$\chi_0$
  to the critical embedding. The blue dotted line represents the
  analytic result obtained by using a step potential (where~$F=0.84$
  was used in~\protect\eqref{e:rindlerQNM}). The values~$\chi_0=0.94$
  and $\chi_0=0.9621$ correspond to the points where the embedding
  becomes metastable and unstable, respectively.\label{f:firstquasi2}}
\end{figure}

This expression is rather appealing. The real part is the one obtained
if one had an IR infinite wall at $\sigma_{\text{step}}$. That is, it
corresponds to the meson masses of the Minkowski embedding if one
identifies $\sigma_{\text{step}}$ with $\sigma_{\text{IR}}$ there.  In
fact, from figures~\ref{f:V_Mink} and~\ref{f:V_bh} we see that
these values are comparable if $z_0$ and $y_0$ are of the same order.
On top of that, there is a small, negative imaginary part (remember
$\sigma_{\text{step}} < 0$).

The numerical analysis of the quasi-normal frequency spectrum away
from the critical embedding, as presented in~\cite{Hoyos:2006gb} for a
scalar excitation, requires some care if one wants to extend it to the
regime of near-critical embeddings. In the coordinates used here, we
found reliable numerical results for the first quasi-normal mode,
which match smoothly onto our analytic results. The result for the
D3/D7 system is shown in
figure~\ref{f:firstquasi2}. Equation~\eqref{e:rindlerQNM} thus
provides us with an endpoint for the vector analog of the spiralling
trajectories of the quasi-normal modes shown in figure~4
of~\cite{Hoyos:2006gb}.

Summarising the present section, we have seen that any finite number
of meson masses in the D3/D7 system come down to a single
value~$\sqrt{V_{\text{crit}}}$. Similarly, any finite number of
quasi-normal modes come down to the same value as the critical
embedding is approached from the black hole phase. Since the masses
and quasi-normal modes are labelled by a single excitation number~$n$,
there is in principle an infinite number of ways to relate the states
in these two spectra, and the connection between mesons and
quasi-normal modes thus seems more subtle than previously
suggested~\cite{Myers:2007we,Erdmenger:2007jab}. Note that the
degeneracy is not in contradiction with the Von~Neumann \& Wigner
theorem~\cite{Neumann:1929a}. In our case, a finite change of the
parameter (here given by~$T/m_q$) leads to an infinite change of the
potential (the logarithmic growth which diverges at the critical
embedding), while the Von~Neumann \& Wigner theorem requires the
potential to change smoothly.

\section{The D4/D8 system}
\label{s:D4D8}
\subsection{Schr\"odinger potentials}
\label{s:D4D8pot}

The D4/D8 embeddings used in the Sakai-Sugimoto model are different
from the cases analysed so far. Instead of being described by a curve
in the~$r-\lambda$ plane, these embeddings are given by curve in the
$u-\tau$ subspace, i.e.~these are more like the ``equatorial
embeddings'' of the D3/D7 system. There is a family of Minkowski-type
embeddings, labelled by the lowest point~$u_0$ of the D8 brane in the
$u$-direction. The black hole embeddings are labelled by the distance
in the~$\tau$ direction between the two halves of the D8~brane, but
this distance is a modulus and the spectrum is independent of it.

In the intermediate-temperature phase, where chiral symmetry is
broken, the vector meson fluctuation equation
reads~\cite{Peeters:2006iu} (we set~$u_T=L=1$ from now on)
\begin{equation}
\partial_u \Big( 
u^{5/2} \gamma^{-1/2} f_4(u)^{1/2} \partial_u \psi_{(n)}\Big) 
+ u^{-1/2}\gamma^{1/2} f_4(u)^{-1/2} \tilde{m}_{(n)}^2 \psi_{(n)} = 0\,.
\end{equation}
where
\begin{equation}
\gamma = \frac{u^8}{u^8 f_4(u) - u_0^8 f_4(u_0)}\,.
\end{equation}
Following the by now familiar procedure we thus construct the
Schr\"odinger radial variable~$\sigma$ through
\begin{equation}
\sigma(u) = \pm \int_{u_0}^{u} \frac{u^4\,{\rm d}u}{\sqrt{u^3 - 1}
\sqrt{u^5 (u^3 -1) - u_0^5 (u_0^3 -1)}} \,.
\end{equation}
We have defined $\sigma=0$ as the tip of the brane ($u=u_0$) and the
two signs correspond to the two branches of the brane worldvolume.
For any $u_0 >1$, it is clear that
$\sigma_{\text{max}}=\sigma(u=\infty)$ is finite, so $\sigma$ lives in
a finite range. However, as $u_0 \to 1$, $\sigma_{\text{max}}$
diverges. The rate of divergence can be computed by making a
coordinate change~$\delta = (u/u_0)^3 - 1$, which leads to
\begin{equation}
\sigma_{\text{max}}(\epsilon) = \frac{1}{3}
\int_0^\infty
\frac{(1+\delta)^{-\frac{1}{6}}}{\sqrt{\delta}{\sqrt{\delta +
      \epsilon}}}{\rm d}\delta\,,\qquad\text{where}\qquad
\epsilon = 1 - u_0^{-3}\,.
\end{equation}
This integral can be evaluated by splitting it into two regions, $0<
\delta < \sqrt{\epsilon}$ and $\sqrt{\epsilon} < \delta <
\infty$. This leads to two integrals which can be evaluated
analytically, up to corrections which vanish as~$\sqrt{\epsilon}$. The
result is
\begin{equation}
\label{e:sigmamax}
\sigma_{\text{max}}(\epsilon) = \frac{1}{6}\Big(\sqrt{3}\pi + 8 \log 2 + \log 3
-2 \log(u_0 - 1)\Big)  + {\cal O}(\sqrt{\epsilon})\,.
\end{equation}
The Schr\"odinger potential in the $u$-variable is:
\begin{equation}
V=\frac{u^5 ( 2 u^6 + 2 u^3 -4) + u_0^5 (u_0^3 -1) (6 u^3 -9)}{4u^{10}} \,.
\end{equation}
Its limiting behaviour at the tip and far away from it is given by
\begin{equation}
V(u=u_0) =  \frac{8u_0^6 -13 u_0^3 +5}{4u_0^5} \,,\qquad
V(u\rightarrow \infty) = \infty\,.
\end{equation}

\begin{figure}[t]
\begin{center}
\includegraphics[width=.32\textwidth]{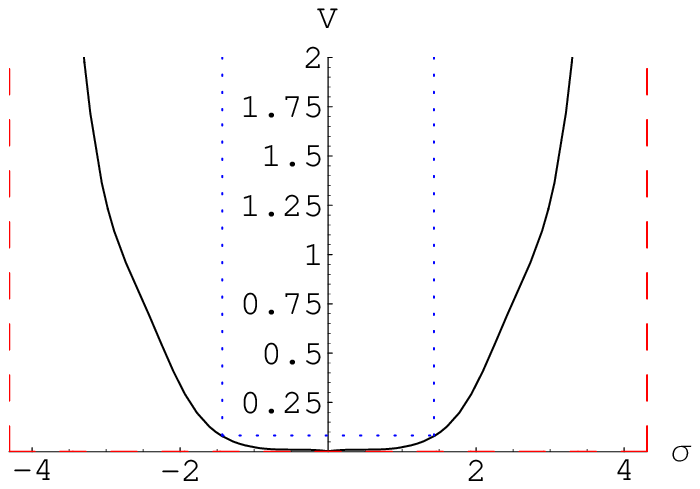}
\includegraphics[width=.32\textwidth]{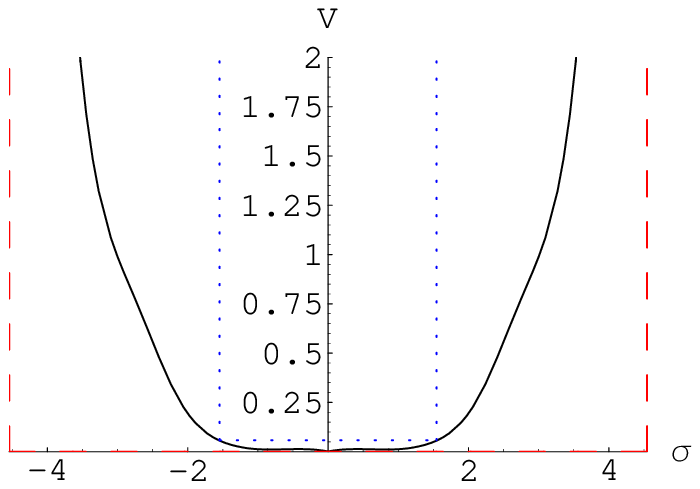}
\includegraphics[width=.32\textwidth]{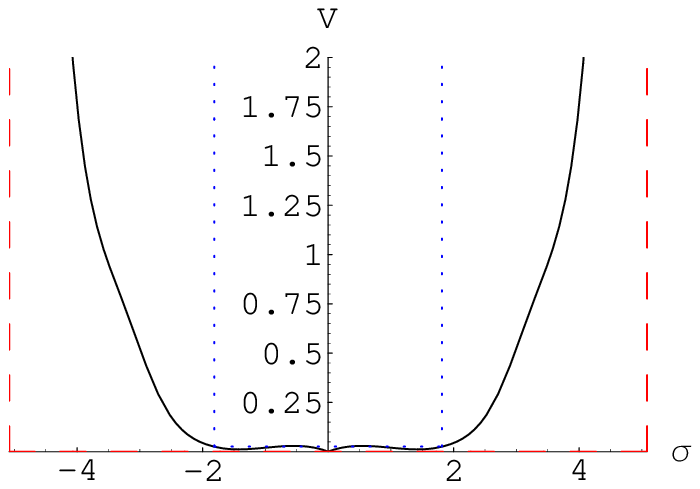}
\end{center}
\caption{The Schr\"odinger potential for the Sakai-Sugimoto system in
  the phase with broken chiral symmetry, for various decreasing values
  of the position~$u_0$ of the tip of the brane. The dashed red lines
  (outer box) denote the potential used for the computation of the
  lower bound, while the dotted black lines (inner box) denote the
  potential for the upper bound (with~$q=2$).\label{f:SSpotentials}}
\end{figure}

The dependence of~$V$ on the~$\sigma$ variable can be computed
numerically. The result for various values of~$u_0$ is plotted in
figure~\ref{f:SSpotentials}. These plots suggest that it should be
possible to find two square well potentials, which provide an upper
resp.~lower bound on the potential. The lower bound is obtained by
using a width which is twice the expression~\eqref{e:sigmamax} (the
red dashed outer box in figure~\ref{f:SSpotentials}).

For the width of the upper bound potential, we pick a point~$\sigma_i$
on the~$\sigma$ axis for which we can show that the potential goes to
zero in the limit~$u_0\rightarrow 1$, and which is such
that~$\sigma_i\rightarrow \infty$ in this limit. A suitable point is
defined by~$\delta = \epsilon^{1/q}$ for some positive~$q$. This point
satisfies
\begin{equation}
\label{e:sigmai}
\sigma_i = \frac{1}{3}
\int_0^{\epsilon^{1/q}} 
\frac{1}{\sqrt{\delta}{\sqrt{\delta+\epsilon}}}{\rm d}\delta = 
\frac{2}{3}\log 2 - \frac{q-1}{3q} \log\epsilon + {\cal
  O}(\epsilon^{\frac{q-1}{q}})\,.
\end{equation}
The potential~$V(\sigma_i)$ behaves as~${\cal O}((u_0-1)^{1/q})$ and
thus goes to zero. For sufficiently small~$u_0$ the potential is
monotonic, and thus we conclude that~$V(\sigma<\sigma_i)$ goes to zero
as well. Furthermore, we can take the formal
limit~$q\rightarrow\infty$ in~\eqref{e:sigmai}. In this limit, the box
size again diverges logarithmically as~$u_0\rightarrow 1$, with the
same coefficient of the logarithm as in~\eqref{e:sigmamax}. We will
use a box with width twice that of~\eqref{e:sigmai}
for~$q\rightarrow\infty$ to obtain an upper bound to the potential (the
black dashed inner box in figure~\ref{f:SSpotentials}).

\begin{figure}[t]
\begin{center}
\includegraphics[width=.45\textwidth]{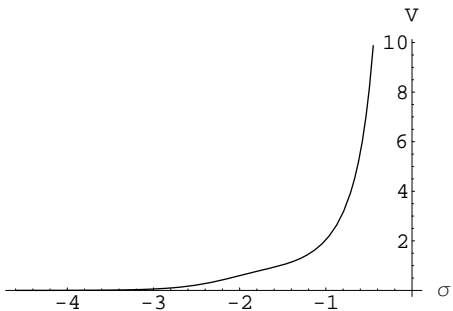}
\end{center}
\caption{The Schr\"odinger potential for the Sakai-Sugimoto system in
  the phase where chiral symmetry is restored. The~$\sigma$
  coordinate spans the half-line~$-\infty < \sigma < 0$.\label{f:SSchisym}}
\end{figure}

In the phase where chiral symmetry is restored, the vector
meson fluctuations are governed by the equation
\begin{equation}
- u^{1/2} (1 - u^{-3} ) \partial_u \Big( u^{5/2} (1-u^{-3}) \partial_u
\psi(u)\Big) = \tilde{\omega}^2 \psi(u)\,.
\end{equation}
A Schr\"odinger equation is obtained by using
\begin{equation}
\Gamma = \frac{1}{u^{1/2} (1-u^{-3})}\,,\qquad
\Sigma = \frac{1}{u^{3/2} (1-u^{-3})}\,,\qquad
\Xi = u^{1/2}\,,\qquad
\tilde{\psi} = \Xi \psi\,.
\end{equation}
The Schr\"odinger coordinate, obtained using~${\rm d}\sigma/{\rm d}u =
\Sigma$, is a complicated function of~$u$ which does not admit a
simple inverse. We therefore restrict to a numerical plot of the
potential in figure~\ref{f:SSchisym}. In the~$u$ coordinate it is
given by
\begin{equation}
\label{e:highTV}
V = \frac{u^6 + u^3 - 2}{2 u^5}\,.
\end{equation}
This potential is quite featureless and we will hence resort to a
numerical determination of the quasi-normal mode spectrum for the
D4/D8 case (this potential is similar to the one for the equatorial
D3/D7 embedding).

\subsection{Mesons and quasi-normal modes}
\label{s:D4D8spectrum}

\begin{figure}
\begin{center}
 \includegraphics[width=0.45\textwidth]{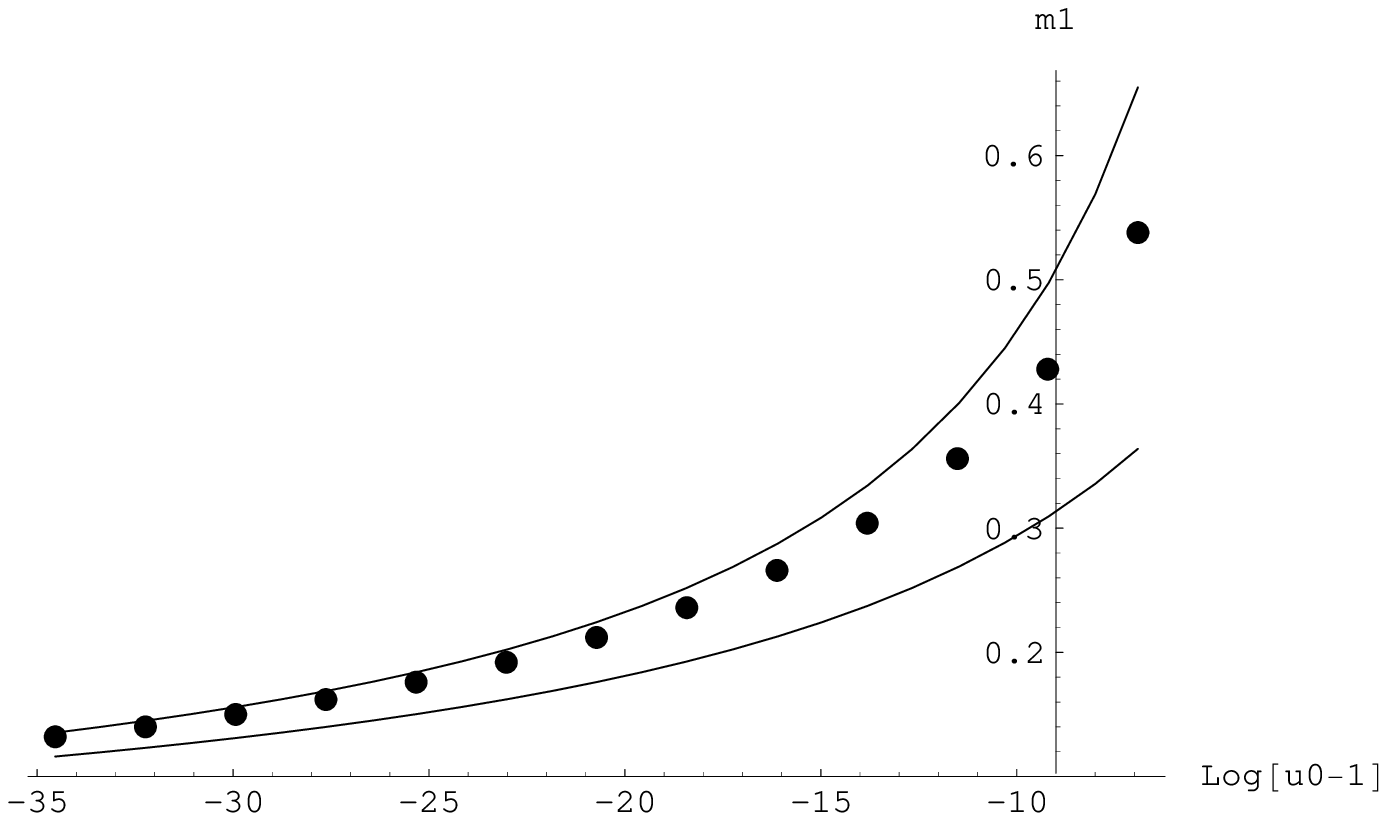}\qquad
\includegraphics[width=0.45\textwidth]{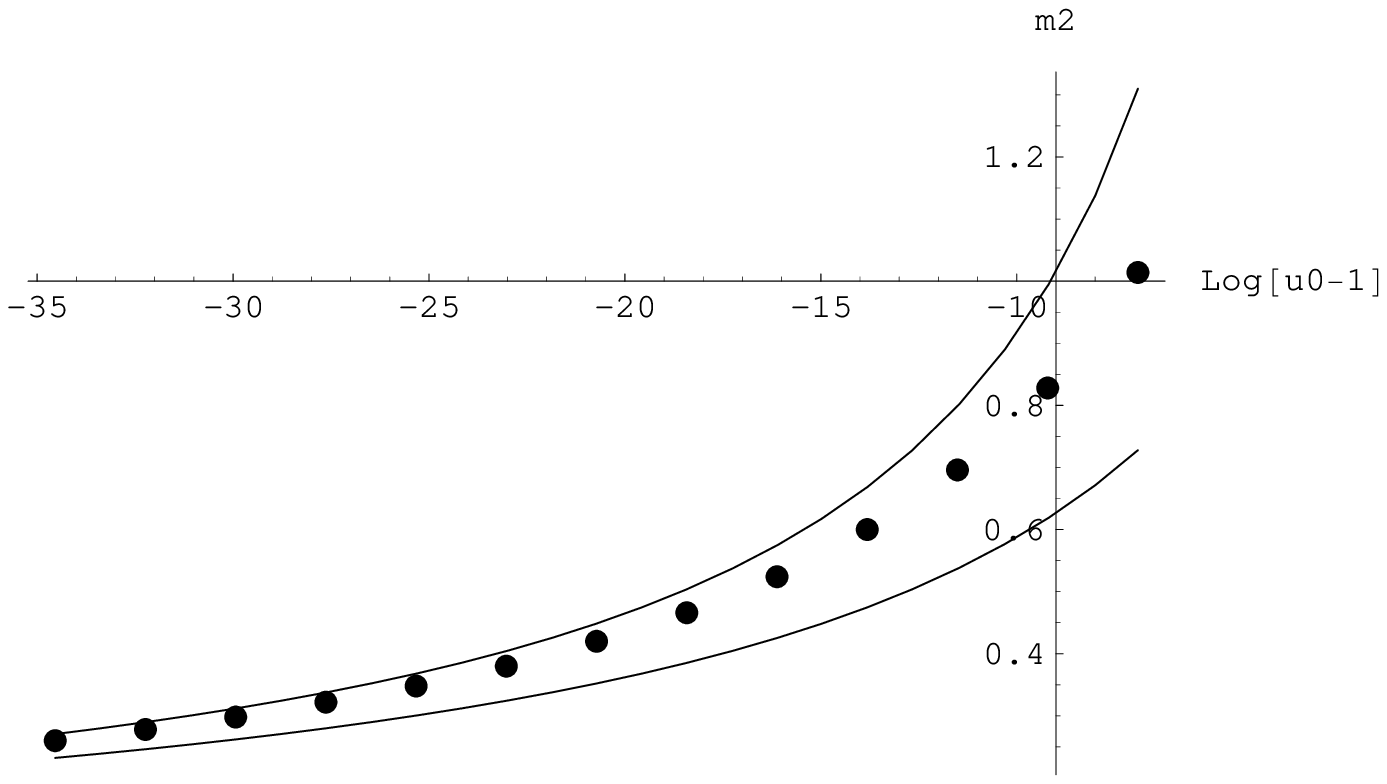}
\end{center}
\caption{The first vector (left) and axial vector (right) meson masses
  computed numerically, versus the analytic upper and lower bounds
  computed~\protect\eqref{e:analytic_upper}
  and~\protect\eqref{e:analytic_lower}.\label{f:figfit}}
\end{figure}

Using the square well potentials which we have shown in the previous
section to be upper resp.~lower bounds for the full Sakai-Sugimoto
potential, it is easy to compute the mass spectrum in
the~$u_0\rightarrow 1$ limit. From the box with
a width given by twice~\eqref{e:sigmamax} we find
\begin{equation}
\label{e:analytic_lower}
\tilde{m}_{(n)} (u_0) \gtrsim \frac{ n \pi}{2 
\left( \frac{\sqrt3 \pi}{6} + \frac43 \log 2 + \frac16 \log 3 -\frac13
\log(u_0 - 1) \right)}\,,\qquad n \in {\mathbb N}\,,
\end{equation}
whereas the box with width given by twice~\eqref{e:sigmai} we get
\begin{equation}
\label{e:analytic_upper}
\tilde{m}_{(n)} (u_0) \lesssim \frac{ n \pi}{2 
\left( \frac{2}{3}\log 2 - \frac{1}{3}\log 3 - \frac{1}{3}\log (u_0-1)
\right)}\,,\qquad n\in{\mathbb N}\,.
\end{equation}
To re-instate the $L$ and $u_T$ dependence, note that
\begin{equation}
\label{e:tildem}
\tilde{m} = \frac{L^{3/2}}{u_T^{1/2}} m = \frac{3m}{4\pi T}\,.
\end{equation}
These expressions both go to zero as~$u_0\rightarrow 1$; see
figure~\ref{f:figfit} for a comparison with the numerically determined
spectrum.

\begin{figure}
\vspace{1ex}
\begin{center}
\includegraphics[width=.4\textwidth]{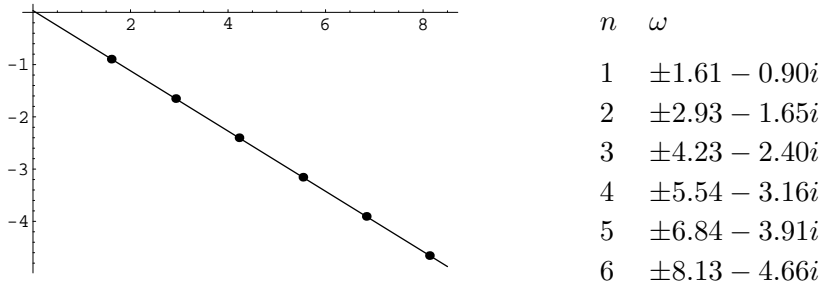}
\qquad\qquad
\raisebox{10ex}{\begin{tabular}{ll}
$n$  & $\omega$ \\[1ex]
1  & $\pm 1.61 - 0.90i$\\
2  & $\pm 2.93 - 1.65i$\\
3  & $\pm 4.23 - 2.40i$\\
4  & $\pm 5.54 - 3.16i$\\
5  & $\pm 6.84 - 3.91i$\\
6  & $\pm 8.13 - 4.66i$
\end{tabular}}
\end{center}
\caption{The spectrum of the first twelve quasi-normal vector modes of the
 chirally symmetric phase of the Sakai-Sugimoto model (each mode is
 twofold degenerate because of chiral symmetry).\label{f:SSqnm}}
\end{figure}

Because the potential in the high-temperature phase~\eqref{e:highTV}
does not take a simple form, we again have to resort to numerics in
order to find the quasi-normal mode spectrum. The first six vector
meson masses are given in figure~\eqref{f:SSqnm} (each of these is
twofold degenerate because of chiral symmetry). A fit to a linear
function yields
\begin{equation}
\tilde{\omega}_n = \pm (0.31 +1.305 n) - (0.145 + 0.752 n)i\,,
\end{equation}
(where~$\tilde{\omega}$ is related to $\omega$ by a relation similar
to~\eqref{e:tildem}).  The numbers agree with the analysis presented
in~\cite{Evans:2008tv}.

Comparing the real parts of the quasi-normal modes with the meson
masses in the intermediate-temperature regime, we see that there is no
room to connect the two in a smooth way. The masses of any finite
number of mesons come down to zero as the critical, overheated
embedding is approached. In contrast to this, the quasi-normal modes
take values which are equally spaced in the complex plane, with a
lower bound on the real part which is non-zero. An identification
along the lines of~\cite{Myers:2007we,Erdmenger:2007jab} thus does not
seem possible for the Sakai-Sugimoto model.

\section{Conclusions and discussion}

We have analysed the form of the Schr\"odinger potential for vector
meson fluctuations in various holographic duals to gauge theories with
matter. These potentials exhibit some generic structure which we
expect to see for more general classes of models. In the Rindler
region, a Minkowski embedding will lead to a box-shaped potential,
while a black hole embedding will lead to a step. When no Rindler
approximation can be made, which is the case for the equatorial
embeddings, more of the background can be seen, and the potential is
typically more complicated.

In the non-equatorial type transition of the D3/D7 system, we have
shown how any finite number of meson masses come down to a fixed
value. Similarly, when the embedding is approached from the black hole
side, any finite number of quasi-normal modes approaches this same
value. Lacking any quantum numbers to distinguish them, the connection
between these spectra is thus more subtle than previously
suggested.\footnote{In~\cite{Myers:2007we} it was observed that peaks
  appear in the spectral function, which is consistent with our
  results. However, our calculations show that these peaks all
  converge on~$\tilde\omega = 0.75$. The spectra are
  symmetric only for the artificial choice of fixed~$z_0/y_0$.}

For the D4/D8 embeddings (``equatorial''), we have shown that all
meson masses come down to zero. This is consistent with the fact that
there is a proper chiral symmetry restoration after the transition, in
sharp contrast to the D3/D7 system. The quasi-normal mode spectrum, on
the other hand, is equally-spaced with a non-zero lowest frequency
mode. This result does not support the idea that quasi-normal modes
are continuously connected to meson modes.
It is conceivable that, if one would introduce a bare quark mass into
the Sakai-Sugimoto model (along the lines
of~\cite{Bergman:2007pm,Dhar:2007bz,Casero:2007ae}), its behaviour
would be more similar to the D3/D7 case. It would also be interesting
to verify if the qualitative results we have obtained for the D4/D8
model are realised in its non-critical version~\cite{Casero:2005se},
whose thermal phase structure was analysed in~\cite{Mazu:2007tp}.

We have not yet touched the issue of chemical potentials. In the
presence of a baryon chemical potential, the spectral function
exhibits, for large enough~$m_q/T$, sharp peaks close to the critical
embedding~\cite{Erdmenger:2007jab}. Moreover, these peaks converge on
the meson masses for the Minkowski embeddings. However, while the
embeddings for large chemical potential are very close to the
Minkowski embeddings (and hence the agreement in the spectrum), they
are not related to each other by a small deformation in the gauge
theory parameter space. The situation is reminiscent of a strong/weak
duality in chemical potential. It would be interesting to understand
this behaviour for isospin potentials as well~\cite{Aharony:2007uu}.

\section*{Acknowledgements}

We thank Ofer Aharony for extensive comments on the draft, and Johanna
Erdmenger, Nick Evans and Andrei Starinets for discussions. The work
of \mbox{AP \& KP} was supported by VIDI grant 016.069.313 from the
Dutch Organisation for Scientific Research (NWO). The work of AP is
also partially supported by EU-RTN network MRTN-CT-2004-005104 and
INTAS contract 03-51-6346.

\appendix
\section{Appendix}
\subsection{Coordinate frames}
\label{a:frames}

We collect here a number of technical details related to the
coordinate choices made in the main text.  We will rewrite the
metric~\eqref{e:bgmetric} in two different ways, as depicted in
figure~\ref{f:setup}. One coordinate frame is better adapted for
computations with Minkowski embeddings, while the other one is more
suitable for black hole embeddings.

The first step is to define coordinates such that the second part of
the metric~\eqref{e:bgmetric} becomes conformally flat. This is
achieved with
\begin{equation}
u = u_T \left( \tilde \rho^{\frac{7-p}{2}} + \frac{1}{4\,\tilde
\rho^{\frac{7-p}{2}}}\right)^{\frac{2}{7-p}}\,,
\quad K_p (\tilde\rho) = \frac{u^\frac{p-3}{2}}{\tilde\rho^2}\,\,.
\end{equation}
The metric~\eqref{e:bgmetric} becomes
\begin{equation}
\label{metricrho1}
{\rm d}s^2 = \left(\frac{u}{L}\right)^{\frac{7-p}{2}}\Big(
\!- f_p {\rm d}t^2 + \delta_{ij} {\rm d}x^i {\rm d}x^j
\Big) + L^\frac{7-p}{2} K_p(\tilde\rho)\Big(d\tilde\rho^2 +
\tilde\rho^2 {\rm d}\Omega_{8-p}^2\Big)\,\,.
\end{equation}
We can reexpress the flat factor inside the last bracket
by defining $\tilde\rho^2 = r^2 + \lambda^2$, and we
get
\begin{equation}
\label{metricrlambda}
{\rm d}s^2 = \left(\frac{u}{L}\right)^{\frac{7-p}{2}}\Big(
\!- f_p {\rm d}t^2 + \delta_{ij} {\rm d}x^i {\rm d}x^j
\Big) + L^\frac{7-p}{2} K_p \Big({\rm d}\lambda^2 +
\lambda^2 {\rm d}\Omega_{6-p}^2 + {\rm d}r^2 +r^2 {\rm d}\phi^2 \Big)\,\,.
\end{equation}
This is the $r-\lambda$ system depicted in figure \ref{f:setup}, and
for $p=4$, it corresponds to the coordinate system defined
in~\cite{Kruczenski:2003uq}. We have found this frame convenient to
numerically analyse the Minkowski embeddings which reach~$\lambda = 0$
at some~$r=r_0$, which parametrises the one-parameter family of
embeddings (Schr\"odinger potentials for various values of~$r_0$ in
the D3/D7 case are plotted in figure~\ref{f:V_Mink}).  The critical
embedding corresponds to~\mbox{$r_0 = 2^\frac{2}{p-7}$}.

We now define a $\rho-\chi$ coordinate system (as shown in figure
\ref{f:setup}) following~\cite{Mateos:2006nu}. We first
rescale the $\rho$ coordinate defined above as:
\begin{equation}
\rho = 2^\frac{2}{7-p} \tilde \rho\,,
\end{equation}
such that the horizon is at $\rho = 1$ and write the metric of the
sphere as ${\rm d}\Omega_{8-p}= {\rm d}\theta^2 + \sin^2 \theta {\rm d}\Omega_{6-p}
+\cos^2 \theta {\rm d}\phi^2$.  If we now define $\chi=\cos \theta$, we get
\begin{multline}
\label{metricrhochi}
{\rm d}s^2 = \left(\frac{u}{L}\right)^{\frac{7-p}{2}}\Big(
\!- f_p {\rm d}t^2 + \delta_{ij} {\rm d}x^i {\rm d}x^j
\Big) \\[1ex]
+ L^\frac{7-p}{2} u^\frac{p-3}{2} \left( \frac{{\rm d}\rho^2}{\rho^2}  +
\frac{{\rm d}\chi^2}{1-\chi^2} + 
(1-\chi^2) {\rm d}\Omega_{6-p}^2 +\chi^2 {\rm d}\phi^2 \right)\,\,,
\end{multline}
where $u$ should be understood to be a function of $\rho$.  This
coordinate system is convenient to study the black hole embeddings,
parametrised by $\chi_0\equiv \chi|_{\rho=1}$ (with $\chi_0=1$ being
the critical embedding).  Some Schr\"odinger potentials for various
values of~$\chi_0$ in the D3/D7 case are shown in figure \ref{f:V_bh}.
In this set of coordinates, Minkowski embeddings are parametrised
by~$\rho_0 \equiv \rho|_{\chi=1}$.  For embeddings near the critical
one, there are simple relations between these parameters and $y_0,
z_0$ as defined in section~\ref{sec: Rindler}. For the D3/D7 case
($p=n=3$), they read:
\begin{equation}
y_0 = \sqrt2 \sqrt{1-\chi_0}\ L\ (1 + {\cal O}(1-\chi_0))\,\,\qquad
z_0 = L\ (\rho_0 -1)\ (1 + {\cal O}(\rho_0-1))
\end{equation}
These relations, together with~$\rho_0 = \sqrt{2} r_0$, were used to
obtain the last equality in equations~\eqref{e:widthestimateMink}
and~\eqref{e:widthestimate}.

\subsection{Quasinormal frequencies in toy potentials}
\label{appendixB}

We have argued in the main text that for the D3/D7 and D4/D6 cases near
the critical embedding, the Schr\"odinger potential presents a plateau
of diverging width in the $\sigma$-coordinate.  At its edges, there
are some IR and UV features which remain of finite size. Since the
plateau is infinitely big as compared to these finite size details,
one expects that these details only affect the quasinormal frequencies
(or meson masses) in a subleading way.  In this appendix, we will
explicitly check this for two toy Schr\"odinger potentials that
approximate the near critical Schr\"odinger potential for D3/D7 black
hole embeddings.

By looking at the plots in figure~\ref{f:V_bh}, one sees that there is
a bump before the plateau and a dip after the plateau, before the
infinite wall. We stress that the size of the bump and dip remain
constant as the plateau grows unbounded when the embedding approaches
the critical one.  Let us, accordingly, consider the following
approximation for the Schr\"odinger potential which we insert in
(\ref{eq schro1}):
\begin{equation}
\begin{aligned}
&V=0 \quad          &&\text{for~} \sigma < \sigma_{\text{step}} + x_1 \,,\\
&V= V_{\text{crit}} + h_1 \quad &&\text{for~} \sigma_{\text{step}} + x_1<\sigma < \sigma_{\text{step}} \,,\\
&V= V_{\text{crit}}       \quad &&\text{for~} \sigma_{\text{step}} <\sigma < x_2 \,,\\
&V = V_{\text{crit}} -h_2 \quad &&\text{for~} x_2 <\sigma < 0 \,,
\end{aligned}
\end{equation}
and an infinite wall at $\sigma =0$. We want to consider
$V_{\text{crit}},h_1,h_2 >0$ and $\sigma_{\text{step}}, x_1, x_2 <0$,
in a limit $\sigma_{\text{step}} \to -\infty$ with
$V_{\text{crit}},h_1,h_2,x_1, x_2$ fixed. We can obtain a solution of
the Schr\"odinger problem with IR incoming boundary conditions
($\tilde\psi = e^{-i \omega \sigma}$ in region 1) as a series in
$\frac{1} {\sigma_{\text{step}}\sqrt{V_{\text{crit}}}}$.  The solution
is the same as in the simple step potential, see (\ref{e:rindlerQNM}),
with $F$ depending on the various parameters,
\begin{multline}
F=-i\
\left(\sqrt{h_1 h_2 V_{\text{crit}}} \cosh (\sqrt{h_1 x_1}) - 
i h_1 \sqrt{h_2} \sinh (\sqrt{h_1 x_1})
\right)^{-1} \\
\times \bigg( \begin{aligned}[t] 
& \cosh (\sqrt{h_1 x_1})\left[ 
\sqrt{h_1 h_2} V_{\text{crit}} x_2 + i \sqrt{h_1 h_2 V_{\text{crit}}}
- \sqrt{h_1} V_{\text{crit}} \tan (\sqrt{h_2} x_2)
\right] \\
+ & 
\sinh (\sqrt{h_1 x1}) \left[
-i h_1 \sqrt{ h_2 V_{\text{crit}}} x_2 - \sqrt{h_2} V_{\text{crit}}
+ i h_1 \sqrt{V_{\text{crit}}} \tan (\sqrt{h_2} x_2)
\right]\bigg)\,.
\end{aligned}
\end{multline}
If we insert $h_1 = h_2 =.1$, $x_1=x_2=-2$, which is approximately
appropriate for the potentials we are dealing with, and also
use~$V_{\text{crit}}=9/16$, we get: $F\approx
.65 -1.72 i$.

One may also worry if the exponential tail of the Schr\"odinger
potentials, which is proportional to $e^{2\sigma}$ as $\sigma \to -\infty$
could affect the computation. Let us address this point by considering
another toy potential slightly more complicated than the 
one in figure \ref{f:approxV}:
\begin{equation}
\begin{aligned}
V &=V_{\text{crit}} e^{2\sigma - 2 \sigma_{\text{step}}}  \quad &&{\textrm { for } } \sigma < \sigma_{\text{step}}  \,,\\
V &= V_{\text{crit}}  \quad &&{\textrm { for } } \sigma_{\text{step}}  < \sigma  < 0 \,,
\end{aligned}
\end{equation}
Again, one can make an expansion for the $\sigma_{\text{step}} \to -\infty$ 
limit and the solution for the quasinormal frequencies is again
(\ref{e:rindlerQNM}) where $F$ now can be written in terms of the
modified Bessel function:
\begin{equation}
F=\frac{-2i\ I_{-i\sqrt{V_{\text{crit}}}}(\sqrt{V_{\text{crit}}})}
{I_{-1-i\sqrt{V_{\text{crit}}}}(\sqrt{V_{\text{crit}}})+I_{1-i\sqrt{V_{\text{crit}}}}(\sqrt{V_{\text{crit}}})}
\end{equation}
If we insert $V_{\text{crit}} = 9/16$, we find $F\approx 1.1 - .32 i$.

These two examples explicitly show how in the $\sigma_{\text{step}}
\to -\infty$ limit, the details of the potential only appear in
$\tilde \omega_{(n)}$ at order $(\sigma_{\text{step}}
\sqrt{V_{\text{crit}}})^{-3}$ (which is however the leading imaginary
part for the quasinormal frequencies). In summary, this shows that
finite details do not change the discussion based on the infinite
plateau.


\begingroup\raggedright\endgroup

\end{document}